\documentstyle[aps,prl,epsf,rotate]{revtex}

\def\be{\begin{equation}}
\def\ee{\end{equation}}
\def\bea{\begin{eqnarray}}
\def\eea{\end{eqnarray}}

\begin{document}

\title{  
Scaling-violation phenomena and fractality in the human posture control systems }

\author{
   Stefan Thurner,$^{1,2}$ 
    Christian Mittermaier,$^{3}$ 
    Rudolf Hanel$^{4}$
    and Klaus Ehrenberger$^{1}$ \\
    $^{1}${\it Klinik f\"ur HNO, Universit\"at Wien, Austria  } \\
    $^{2}${\it Institut f\"ur Mathematik, NuHAG, Universit\"at Wien, Austria }\\
    $^{3}${\it Klinik f\"ur Physikalische Medizin, Universit\"at Wien, Austria  } \\
    $^{4}${\it Institut f\"ur Theoretische Physik, Universit\"at Wien, Austria }\\
}

\maketitle
\begin{abstract}
By analyzing the movements of quiet standing persons by means 
of wavelet statistics, we observe multiple scaling regions in the 
underlying body dynamics. The use of the wavelet-variance function 
opens the possibility to relate scaling violations to different modes of 
posture control. We show that scaling behavior becomes 
close to perfect, when correctional  movements are dominated by the 
vestibular system.

\vspace{0.5cm} 


\vspace{0.5cm} 

  
\end{abstract}

\section{Introduction}
Over the last years evidence has been compiled that many physiological 
systems and processes exhibit scaling laws in the corresponding time 
series \cite{BAS94}, due to the existence of control  mechanisms with 
a minimum degree of complexity. Irregularities in physiological 
processes may lead to scaling violations which recently  have 
successfully been used  to detect abnormalities of the underlying 
biological systems such as the human heart \cite{THU981,THU982}.  

For some reason nature choose to let man walk upright. 
This has led, over time,  to a highly complex balance system, which is not 
based on one single physical principle alone. 
Balance of a normal human being is 
the result of a most likely nonlinear \cite{NEW93} 
superposition of sensory signals originating from the 
vestibular, visual and somatosensory  systems, which govern  
a cascade of muscular correctional movements in order to maintain an 
upright position, for an overview see e.g. \cite{MASS94}. 
The existence of three different mechanisms for maintaining balance 
provides the body with  backup systems. If the vestibular system is severely 
damaged or not fully functioning, visual and somatosensory  information 
will gain importance and  take up for the loss.  

Human posture can be quantified by measuring the displacement of the 
center of pressure under the feet of a quiet standing person, 
leading to time-varying trajectories Fig. 1.  
It is known in posturography that in case of damages of one 
component of the balance system certain characteristic frequency 
bands will change their relative weights, see e.g. \cite{KOHEN1}. 
Frequency bands provide a crude indicative of typical functional aberrations in the 
visual, vestibular,  somatosensory and central nervous system. 
The visual system is linked to frequencies in the range of 
0-0.1 Hz which dominate normal steady and undisturbed posture. 
Vestibular stress and disturbances will lead to enhanced frequencies 
in the 0.1-0.5 Hz band; frequencies between  0.5 and 1 Hz indicate 
somatosensory activity and postural reflexes mediated by the lower extremities. 
Finally, signs for malfunctions in the central nervous system usually correlate 
to high frequencies, i.e. 1 Hz and higher \cite{KOHEN1,KOHEN2}. 

In this paper we provide first evidence that the different control mechanisms,  
at work at their characteristic time scales, lead to different 
scaling behavior in the  center of pressure trajectories. 
By using a novel sensitive scaling measure based on wavelet statistics, 
we not only show  that quiet standing is a 
correlated noise which has been observed before \cite{JJC94,JJC95}  
but that in normal posture the mode of the control mechanism 
at work, is directly related to one of the multiple scaling regions 
with their characteristic scaling exponents. 
Scaling is found to be close to perfect (one single scaling exponent) 
in the case where visual and 
tactile senses are  excluded, and the vestibular system plays the dominant  
role. This difference in scaling behavior could  be used to directly quantify 
the relative importance of the vestibular system in relation to the entire 
human balance system.     
The wavelet measure is able to capture this 
scaling differences from standard posturography data on an individual basis.  
This suggests that the measure could be of clinical use in diagnostic of the 
functional quality of the different control mechanisms of human posture.  

The paper is organized as follows: In Sect. II we review the concept of scaling 
statistics and conventional  methods for its quantitative treatment. We further discuss  
a novel wavelet-variance function method. In Sect. III the experimental setup and 
the way of data acquisition are described. Results and a comparison of the 
different methods are presented in Sect. IV.  
Section V concludes with a discussion of the findings 
and their potential clinical usefulness. 

\section{Methods}

To completely characterize processes from a stochastic point of view, the knowledge of all 
possible joint probabilities of the various events occurring in these
processes is necessary. 
Fractal stochastic processes exhibit scaling in their statistics,  which  
naturally lead to power-law behavior. Consider a statistic $f$ 
which depends continuously on the scale $x$ over which measurements are taken. 
Suppose the scale gets changed by a factor $a$ and induces the statistics 
to get scaled by another factor $g(a)$, related to the factor but independent 
of the original scale: 
\begin{equation}
f(ax)=g(a)f(x) \,    . 
\end{equation}
The only nontrivial solution to this scaling equation for real valued functions 
and arguments is 
\begin{equation}
f(x)=bg(x)    \quad {\rm with} \quad g(x)=x^c  \,    ,
\end{equation}
for some constants $b$ and $c$, e.g.  \cite{steve,THU973}. The particular case of 
a fixed $a$ admits a more general solution,
$g(x;a)=x^c \, {\rm cos} [2\pi {\rm ln}(x)/{\rm ln}(a) ]$ \cite{scal}. 
 
\subsection{Scaling Measures} 

Over the years there have been developed a vast number of statistics to  
characterize stochastic processes, 
some  which have already been applied to human posturography data.   
Maybe the most direct approach to quantify scaling is to use {\it two-point 
correlation functions}  of the underlying (continuous or discrete) process 
$x(t)$  (variance of $\tau$-increments): 
\begin{equation} 
K(\tau)=\langle [ x(t+\tau) - x(t)]^2 \rangle_t 
\end{equation}  
where $\langle.\rangle_t$ denotes the average over $t$. 
For processes of the kind of fractional Brownian motion the correlation 
$K(\tau)$ scales 
as $K(\tau)\sim \tau^{2C}$. For $C=1/2$ the diffusion result is obtained 
(Brownian motion).  This measure has been used recently to 
show that the dynamical process of quiet standing contains 
long term correlations, and 
thus can be interpreted as a  correlated noise \cite{JJC93,JJC94,JJC95}. 

Another method frequently used to quantify scaling behavior 
of  time series is to analyze the corresponding 
(discrete) {\it Fourier spectra}, 
\begin{equation} 
\tilde X(\omega)= | \sum_{k=1}^{N} x(k) e^{i 2 \pi (\omega-1)(n-1)/N }  |^2  
\propto \omega ^{S}   \,    , 
\end{equation}  
to determine the Fourier-scaling exponent $S$. Fourier analysis strongly depends on 
the stationarity of the underlying signals, which is often not the case for 
real world data. In particular posturography data contains drifts 
originating from slow correctional movements.  
It is common practice in time series analysis to apply  
Fourier methods after some sort  of `detrending' of the signals.

In the course of improving correlation measures, 
factors like the {\it Fano} and {\it Allan factor} have been proposed and applied widely
for stochastic point processes \cite{steve}. 
Those factors are not necessarily limited to 
point processes, and can be used to obtain statistical information of  
time series. 
For an overview, especially for the relations of these factors to the 
power-spectral density, see \cite{THU973}. 
The basic idea behind those and related measures is to compare sets of 
aggregated (over some time interval) events, rather than just looking at 
increment statistics as in the 
case of two-point correlation functions.  
As a further generalization of this 
the concept,  the wavelet-variance function (WVF) has been 
introduced \cite{THU981,THU982}.

\subsection{Wavelet-variance function} 
Multiresolution wavelet analysis \cite{DOBE92,MALL89,MEYE86,ALDR96,ARNE95} 
has proved to be a mathematically clear and practical tool for analyzing 
signals at multiple scales, even 
in the presence of nonstationarities \cite{ABRY96,TEIC96}, which  are 
obviously present in the center of pressure time series $x$ and $y$, Fig. 1b. 
It was shown  that wavelet statistics can be used to 
reveal scaling phenomena in natural systems in a more transparent way 
\cite{THU981,GOLD2,THU982} than 
other methods frequently used. 

In a first step we transform our time series into a (two dimensional) space of 
wavelet coefficients. 
Technically the coefficients are obtained 
by carrying out the discrete wavelet transform (DWT) \cite{DOBE92} of $x$: 
\begin{equation}
W^{{\rm wav}}_{m,n} (x)= 2^{-m/2}  \int_{-\infty }^{\infty} x(t) \psi (2^{-m} t- n) dt
\, ,
\end{equation}
(same for $y$) 
where the scale variable $m$ and the translation variable  $n$ are
integers. The discrete wavelet transform is evaluated at the points $(m,n)$ 
in the scale--interval-number plane.
Smaller scales correspond to more rapid variations and therefore to 
higher frequencies. 

We have carried out this transformation using  a broad range of  
orthonormal, compactly supported analyzing wavelets. Throughout the paper   
we present results for the  
Daubechies 10-tap wavelet. Similar results were obtained using Mallat and Haar wavelets.  
Orthogonality in the DWT provides  that the information represented  at a certain
scale $m$ is disjoint from the information at other scales.
Because certain wavelets  $\psi$ 
have vanishing moments, polynomial trends in the signal 
are automatically eliminated in the process of wavelet transformation 
\cite{ARNE95,TEIC96,ABRY96}. 
This is salutatory in  the case of the time series $x$ and $y$, 
as is evident from the trends apparent in Fig. 1b.  
Since the signal $x(t)$ fluctuates in time, so too does the 
sequence  of wavelet coefficients at any given 
scale, though its mean is zero since   
$\int_{-\infty}^{\infty} \psi(t) \, dt=0$.

In \cite{THU981} it was suggested to investigate the scaling behavior 
of the {\it statistics} of the wavelet coefficients at a particular scale 
of the signal of interest, resulting in the wavelet-variance  
function  
\begin{equation}
\sigma_{{\rm wav}}^2(m) =  \frac{1}{N-1} 
                \sum_{n=1}^N (W^{{\rm wav}}_{m,n}(x) - 
                \langle W^{{\rm wav}}_{m,n}(x) \rangle_n )^2 
    \propto m^{2\Delta} \, ,
\end{equation}
where  $N$ is the number of wavelet coefficients at a given scale $m$ 
($N=M/2^m$, with $M$ being the total number of sample points in the signal).
$\langle W^{{\rm wav}}_{m,n}(x) \rangle_n$ denotes  the mean taken at a given 
scale. For reasonably long signals this quantity will be close to zero  
and can be neglected for practical purposes. 

It is not straight forward to relate the wavelet-variance scaling exponent 
$\Delta$ to the Fourier-scaling coefficient $S$. The exponent $\Delta$ may 
to some extend depend on the choice of the type of wavelets.  However, 
the Allan factor when multiplied by the mean,  
can be seen as a special case of the WVF if the Haar wavelet is taken, 
and a relation to the Fourier-scaling exponent can be established \cite{THU973}. 
Later in this work we present an empirical relation based on fractal Gaussian noise 
(FGN) surrogate data and on the Daubechies 10-tap wavelets. 

\section{Data and experimental setup } 
For measuring the movements of an upright standing person,  
the person was placed on a standard posturography plate \cite{data}, 
which is sensitive to weight shifts. 
Data was gathered  
on 37 healthy  subjects (Age: $31.7 \pm 6.8 $ years, Weight: $66.3 \pm 12.2$ kg, 
Height: $173.0 \pm 8.9$ cm, Sex: 19 females, 18 males), 
who had no history of dizziness,  
have never suffered any leg injuries, 
nor were taking any form of medication before the measurements. 
The sample contains no subject with musculo-skeletal, neurologic or  
vestibular disorder. 
The measurements were taken  
with the subjects looking in the forward direction,  wearing no  shoes 
and no long pants or skirts, to avoid  tactile feedback loops along the legs. 
Individual tests lasted for 20 seconds as is a typical clinical posturography standard, 
points were  sampled at a rate of  50 Hz. 
The reason for keeping the measurements short is to reduce the probability 
that the subject would  
change its standing strategy (toe or hip strategy) within a single measurement.  
Each subject was measured three times to ensure consistency. 

The output of the plate  
is the trajectory of the center of gravity projected onto the $xy$ plane, 
Fig. 1a. 
The actual measured $x$ and $y$ trajectories have been shifted by their mean, 
$ x=x_{\rm measured} -\langle x_{\rm measured} \rangle _t $ (same for $y$). 
For the complete knowledge of a dynamical system  the time series 
of the momenta are also needed. We obtain an estimate of these 
by  taking time derivatives 
(first differences) of the $x$ and $y$ trajectories and denote them 
by $v_x=\frac{d}{dt} \, x$ and $v_y=\frac{d}{dt} \, y$, respectively, 
Fig. 1c. All further analysis has been  carried out on these position and `velocity'  
trajectories. 
Figure 1d shows the phase-space plot of the y-component for the same data. 

To estimate systematic measurement errors, we placed a mass of 75 kg 
on the platform. The resulting trajectories, which relate to vibrations of the 
floor and the measuring equipment, were confined to elongations of less than 0.02 cm. 

\subsection{Experiments}

To be able to study different modes of posture control we performed four types 
of measurements. In the first type the subject was asked 
to stand still with  eyes open (eo). The resulting movements are the bodily 
responses to a mixture of visual, somatosensory and vestibular input. 

In the second test the subjects were asked to close their eyes (ec), 
and were additionally 
provided  darkened swimming glasses, so that no visual input could 
influence posture control. 

The third and fourth test were done 
on a foot plate that could sway in the forward-backward ($y$) direction. 
We refer to this as the eyes-open--moving plate (eomp) 
and the eyes-closed--moving plate (ecmp) tests. 
The experimental design of these tests was to 
reduce somatosensoric effects originating in pressure sensors of the foot. 
Note that in this setup the  pressure on toes and heel is kept  
constant during the measurement due to the moving plate: 
Whenever force is applied by the toes the 
plate moves down in the front, as it will move down in the back whenever 
force is exerted by the heel. 
While in the eomp test,  visual and vestibular systems are active, 
the resulting movements in the $y$-direction  
in the ecmp test will be dominated by the vestibular system alone.    

\section{Results}

In order to relate the wavelet-scaling exponent $\Delta$ to the more familiar 
Fourier-scaling exponent $S$, we generated  
fractal Gaussian noise (FGN) time series, of definite Fourier-scaling exponent 
$\alpha_{\rm theory}$. From those time series we computed the Fourier- scaling 
exponent $S$, which should be the same as $\alpha_{\rm theory}$, 
and the wavelet-scaling exponent $\Delta$. The results for the Daubechies 
10-tap wavelets are given  in Tab. I. 
A linear fit yields an estimate of the  relation of the two scaling measures:
\begin{equation}  
\Delta=-(0.019+ 0.311\,S)  \sim - S/3 \,  . 
\end{equation}  
For each $\alpha_{\rm theory}$ we  generated 50 FGN time series, and averaged 
over the extracted exponents.  

Figures 2a and 2b show $\sigma_{\rm wav}(x)$ and $\sigma_{\rm wav}(y)$ 
as a function of scale $m$ for the  eo and ecmp 
tests of a representative subject.  
Time scales $m$  correspond to data segments of a length of 
$\frac{2^m}{50}$ seconds 
\footnote{The smallest reasonable scale for the given sampling rate of 50 Hz
is thus  $m=1$ or 1/25 sec. The reason why we did not consider smaller scales is most clearly 
seen for the (discrete) Haar wavelet: at scale $m=1$,  only  
two sample points will be in the support of the  wavelet. 
At a smaller time scale ($m=0$),  $\sigma^2_{wav}(0)$ would be 
identical to the variance of the signal.  
 }.  
It is seen that for the eo-case the curve 
is a straight line from scale two on. For the small scales, the slope 
- the wavelet-scaling exponent - denoted by $\Delta_{S}(x)$ 
(scale 1 to 2), 
is clearly less than the slopes at larger  scales $\Delta_L(x)$, 
indicating the onset of white noise 
\footnote{a slope of zero is equivalent to white noise}
in the high frequency region,  i.e. the small scale region. 
For the ecmp situation, no such scaling violation is observed, 
and $\Delta_S \sim \Delta_L$. 
We have checked that this finding is independent of whether discrete or 
continuous wavelets are used, and that it is reasonable to extract the 
scaling exponent $\Delta_S$ from  the smallest 2 scales only in the discrete case. 
For very large (VL) scales ($m=6-10$) we find a significant decrease in the 
corresponding scaling exponent $\Delta_{VL}$ compared to $\Delta_S$ and $\Delta_L$. 
However at those large scales  for our relatively short data sets ($M=1000$)    
statistics becomes sparse  and the estimates for $\sigma^2_{wav}(m>5)$ rather 
unreliable.

These observations lead us to consider an index quantifying the 
degree of scaling violation, determined by the quotient  of the 
wavelet-scaling exponents in large ($\Delta_L$) and small 
($\Delta_S$) scale  regions: 
$\Delta_{\rm sv}= \Delta_L / \Delta_S  $. 
In Fig. 3a the mean values over the whole sample are shown for 
all of the measurements (37 subjects $\grave {\rm a}$  3 times).  
It is clearly observable that the scaling violation measure 
$\Delta_{\rm sv}$ approaches 1 when suppressing visual and tactile control 
(ecmp), and that it nicely distinguishes between the modes of posture control. 
In Fig. 4a  we present $\Delta_{\rm sv}(y)$ for all subjects  
for the eo and ecmp-tests. 
The reduction of $\Delta_{\rm sv}$ from the eo-case 
to the  ecmp-case is apparent. This effect is  
less pronounced in the $x$-direction in which the plate is stable. 
The scatter plot in the  velocity-scaling--position-scaling-violation 
plane also suggests a slight positive correlation of  
velocity-scaling with the index of scaling violation. 

We found that the two-point  correlation function used on the same  
data set shows scaling differences similar to those reported in 
\cite{JJC94,JJC95}, but that it  is not 
sensitive enough to clearly distinguish between different modes of 
posture control. For comparison we show the quotient of 
the corresponding high ($C_{High}$) and low ($C_{Low}$) scaling regions  
$C_{High}/C_{Low}(y)$ in Fig. 3b for the same  measurements. 
We were able to reproduce within errors the two-point correlation 
scaling values in the high and medium frequency domain of \cite{JJC94}. 
Our results are $C_{Low}(y)_{eo}=0.71 \pm 0.07 $, and   $C_{High}(y)_{eo}=0.20 \pm 0.19 $. 
Also the critical time where the scaling regimes of $C$ change could be 
confirmed to be somewhat less than a second, which corresponds to the change 
of wavelet-scaling regimes $\Delta_L(y)_{eo}=0.85 \pm 0.1$ 
to $\Delta_{VL}(y)_{eo}=0.25 \pm 0.27$ at scale $m=6$ (1.28 sec), see Tab. II. 
As for the quotient $C_{High}/C_{Low}$ also $\Delta_{VL} / \Delta_{L} $
is found to be not well suited for a separation of the standing modes, 
especially not on an individual basis. 
Note here that the wavelet-scaling 
regimes from $\Delta_S$ to $\Delta_L$, which are sensitive to posture control, 
change at scales $m=2$ and $3$, 
corresponding to about $1/10^{\rm th}$ of a second.  

Our standard deviation of $C_{High}(y)$  is considerably  larger than reported in \cite{JJC94}, 
which is explained by the fact that the  data length there is   
4.5 times larger (90 sec) than the data used here. 
For our relatively short data segments it was not possible to reliably estimate 
the third scaling region for time scales larger than 10 seconds for all of the 
described methods. In this region scaling vanishes due to the limited  
extend of motion \cite{JJC95}. 

Values for the various scaling exponents and the scaling violation 
quotients  are gathered in Tab. II together with the  
kurtosis and skewness of the increment processes $v(y)$. The latter indicate 
almost Gaussian distributions for the steady plate case which changes 
clearly towards leptocurtic distributions  in the moving-plate scenario.  
The leptocurticity reflects the existence of periods of fast correctional movements 
followed by periods of relatively quiet standing in the moving plate scenarios. 
In financial time series analysis this behavior in dynamics is referred to as 
volatility clustering.

We continue by analyzing scaling behavior of the increment (`velocity') 
processes.   
In Figs. 2c and 2d,  $\sigma_{\rm wav}(v_x)$ and $\sigma_{\rm wav}(v_y)$ 
demonstrate that in the eo-case the wavelet-scaling exponents 
are generally small for the corresponding 
velocities, which means  little  correlation or structure. 
For the  ecmp-case  the wavelet exponent associated to $v_y$, $\Delta_L(v_y)$,  is  
drastically enhanced, while  $\sigma_{\rm wav}(v_x)$ is compatible 
with the eo-case, as expected. 
This behavior was  encountered in all subjects of the sample of 37 healthy 
test persons, as can be inferred from Fig. 4a. 

To compare the wavelet-variance  method to conventional Fourier methods, 
we extracted Fourier-scaling exponents $S$ from the data. 
Since the original signals $x$ and $y$ are 
nonstationary, doubts on the reliability on a straight forward 
use of power spectra are  justified. However, already naive 
Fourier spectra, after windowing,  show typical $1/f$ behavior and it is  
possible to  observe different scaling exponents in most subjects 
in the low ($S_{Low}(y)$) and high frequency regions ($S_{High}(y)$). 
The data does not allow for a stable estimate of the 
degree of the corresponding scaling violation, 
since the inter- and intra-subject variance  are  
high, which is reflected in the large standard deviations in 
the corresponding scaling violation quotient $S_{Low}(y)/S_{High}(y)$ in Tab. II.  
The Fourier-scaling exponents for $S_{Low}(y)$ and 
$S_{High}(y)$ were obtained from fits in the power spectrum in the  region of 
1.45 Hz - 12.45 Hz and 12.45 Hz - 25 Hz respectively.  
We note here that within errors the relation given in eq. (7) still holds approximately 
for the experimental data $\Delta_S$ vs. $S_{High}$ and $\Delta_L$ vs. $S_{Low}$ 
(large scales correspond to low frequencies). 
This could be a sign that the two different scaling regions could be 
successfully modeled with FGN models. However since errors are large for the estimates 
of $S_{High}$ and $S_{Low}$, one should be careful in interpreting this result in terms 
of the nature of the underlying processes at the different scaling regions.  

No problem with nonstationarity will arise when 
the derivatives $v_x$ and $v_y$  are used, and 
clearly $1/f$ behavior is observed. The corresponding Fourier-scaling 
exponents $S(v_y)$ for the individual subjects for two tests 
are gathered in Fig. 4b. 
It is clearly seen that the two tests separate in this variable, 
$S$ drops by a factor of 3 on average.  
In both tests a slight decrease of the scaling exponent 
on age is observable. 

We have computed cross correlation functions of $x$,  $y$, $v_x$ and $v_y$ 
components. 
We could not find a significant dependence of one on the other for all measurements, 
and could such exclude torque effects in quiet standing in healthy subjects.

Additionally we have computed the boxcounting dimension $ D_{\rm bc}$ of the 
graphs  in Fig. 1a for all subjects. The largest box was taken 
to be $6 \times 6$ cm for all measurements. 
The results are given in Fig. 4c. It is remarkable that  in 
this variable, which is a measure of the {\it static} geometry of the 
posture trajectory, separation of the different tests occurs, Tab. II.

\section{Discussion}

All the conventional scaling measures used in this work were able to 
confirm that human posture is indeed a correlated process, and indicate 
that different scaling regions are present (multifractal). 
The wavelet-variance function method proves to exhibit these scaling violations 
in a very precise way so that it becomes 
possible to relate the different control mechanisms to  different scaling exponents 
on an individual basis.  
We show that this is even possible on very short (standard) posturography data, 
clearly outperforming the other methods.  

In particular we demonstrated that dominance of  the vestibular system 
shows close to perfect scaling phenomena in the position time series. 
When  visual and tactile input information 
are superimposed on the vestibular system, scaling violations occur, 
pointing of course at a nonlinear interplay of the underlying systems and  their 
relevant characteristic time scales. 
For the `velocity' time series the effect is even more pronounced 
since the pure vestibular system shows clear scaling, which breaks 
down almost completely for the superposition cases. 

The findings presented here might bear a potential for practical use: 
since the scaling exponents and scaling violation measures can be considered 
as tools which 
measure the relative importance of the vestibular system in comparison with  
the visual and somatosensory systems, 
it seems sensible that they provide a key for a {\it direct} measurement of the 
intactness of the vestibular system. Clinical practice today is to measure 
this intactness of the vestibular system by thermally  disturbing it  and 
measuring  the resulting eye movements.  
A particular interesting subject will be an analysis, along the lines presented here, 
of sway data of patients with dizziness, which often goes hand in hand with 
vestibular stress or malfunction. 

We found first evidence that some scaling measures are age dependent. 
It might be possible to quantitatively relate a loss of complexity in the posture control 
systems of the aging human to changes in scaling measures. 


\newpage 

\begin{table} 
\begin{tabular}{ ccc } 
 $\alpha_{\rm theory}$  &  $-S$     & $\Delta$       \\
\hline 
    0.5 & 0.51(0.06) & 0.14(0.02)  \\
    1.0 & 0.96(0.07) & 0.27(0.03)  \\
    1.5 & 1.50(0.07) & 0.45(0.02)  \\
    2.0 & 2.00(0.06) & 0.64(0.03)  \\
    2.5 & 2.52(0.06) & 0.73(0.04)  \\
\end{tabular}
\caption{Relation of Fourier-scaling exponent and the wavelet-variance exponent
obtained from fractal Gaussian noise surrogate data of definite Fourier-scaling 
exponent $\alpha_{\rm theory}$. To match experimental data the noise sequences 
contained 1000 samples each, 
averages of the exponents were taken over 50 sequences per $\alpha_{\rm theory}$. 
The numbers in brackets are standard deviations.   } 
\end{table}

\begin{table} 
\begin{tabular}{ lcccc } 
                            & eo      & ec       & eomp    & ecmp     \\
\hline 
$ \Delta_L/\Delta_S(y)$     & 1.55(0.73) & 1.46(0.66)  & 1.17(0.46) & 1.02(0.41)      \\
$ \Delta_S(y)$              & 0.55(0.19) & 0.62(0.21)  & 0.76(0.22) & 0.91(0.26)      \\    
$ \Delta_L(y)$              & 0.85(0.10) & 0.91(0.10)  & 0.89(0.10) & 0.92(0.11)      \\    
$ \Delta_{VL}(y)$           & 0.25(0.27) & 0.22(0.26)  & 0.28(0.25) & 0.34(0.28)      \\    
$C_{Low}/C_{High}(y)$       & 3.67(4.00) & 4.02(8.96)  & 3.07(2.00) & 3.34(3.26)      \\
$C_{Low}(y)$                &0.71(0.07)  &0.75(0.07)   &0.71(0.06)  &0.70(0.07)       \\
$C_{High}(y)$               &0.20(0.19)  &0.19(0.25)   &0.23(0.16)  &0.21(0.18)       \\
$S_{Low}(y)/S_{High}(y)$    & 4.97(8.42) & 4.20(5.57)  & 2.35(1.84) & 1.46(0.86)      \\
$S_{Low}(y)$                &-2.55(0.52) &-3.05(0.57)  &-3.42(0.49) &-3.85(0.42)      \\
$S_{High}(y)$               &-0.51(0.77) &-0.73(0.83)  &-1.46(0.94) &-2.64(1.28)      \\
$S(y)$                      &-2.06(0.21) &-2.24(0.30)  &-2.35(0.38) &-2.59(0.50)      \\
$S(d/dt \, y)$              &-0.43(0.23) &-0.74(0.28)  &-1.08(0.28) &-1.50(0.21)      \\   
$D_{\rm bc}$                & 0.67(0.16) & 0.79(0.17)  & 1.00(0.18) & 1.35(0.14)      \\   
kurtosis$(d/dt\, y)$        & 3.49(1.67) & 3.70(0.97)  & 5.34(2.42) & 4.91(2.25)      \\
skewness$(d/dt\, y)$        &0.040(0.198)&0.011(0.225) &0.060(0.424)&0.036(0.394)     \\
\end{tabular}
\caption{Various scaling measures and kurtosis and skewness of the increment processes 
for the four different tests performed (eyes open (eo), eyes closed (ec),
eyes-open--moving plate (eomp) and eyes-closed--moving plate (ecmp)). 
The value in bracket is the standard deviation over all measurements (3 per subject).  
Fits for $S(y)$ and $S(d/dt \, y)$ have been taken over the whole frequency range.}  
\end{table}

\newpage 

\begin{figure}
\caption{Movement of the center of gravity projected on the $xy$-plane (a) 
for open eyes, stable plate. 
The time evolution of the motion for the $x$ and $y$ components is shown 
in (b). The time derivatives  represent momentary `velocity' (c).   
Whereas  positions are clearly nonstationary for the 20 seconds time intervals, 
velocities are.  
(d) Phase-space diagram for the $y$-component.   
The form of the phase-space trajectory reminds on the Duffing oscillator with 
randomized phases. For clarity the plot has been smoothed by moving 
averages of block-size 4. 
} 
\end{figure} 

\begin{figure}
\caption{Wavelet scaling exponents for the eyes open (eo), (left column) and the 
eyes-closed--moving plate (ecmp), (right) test. 
For the position data $x$ and $y$ (top line) 
scaling violation in eo is obvious for low scales.
This violation is absent in ecmp. 
In the velocity data $v_x$ and $v_y$ scaling is generally low in eo, but 
becomes strong in the $y$ component of ecmp.  
 } 
\end{figure} 

\begin{figure}
\caption{Comparison of the wavelet-scaling-invariance   $\Delta_L/\Delta_S(y)$ 
measure (a) and the corresponding correlation quotient $C_{High}/C_{Low}(y)$ (b).
In the latter there is no difference detectable in the mode of posture control.  
eo, ec, eomp and ecmp refrere to the eyes open, eyes closed, eyes-open--moving plate 
and eyes-closed--moving plate tests respectively. 
The errorbars indicate mean standard errors.  
 } 
\end{figure} 

\begin{figure}
\caption{Scaling measures for the  whole sample of 37 healthy persons. 
(a) velocity-scaling--position-scaling-violation plot. 
The scaling-violation measure $\Delta_L/\Delta_S(y)$ is grouped around 
one for the eyes-closed--moving (ecmp) case, while it is larger for the eyes open test (eo). 
The wavelet-scaling exponent 
in $v_y$ is significantly larger for the ecmp than for eo. 
Moreover a  slight positive correlation of velocity-scaling  with 
position-scaling-violation  is observable.
(b) Fourier-scaling exponents for velocities for all individuals ordered by age 
(left - young, right - old).  
(d) Box counting dimension $D_{\rm bc}$ for the same individuals. 
The symbols denote the means of the three identical measurements, the errorbars
are the corresponding mean standard errors.  
} 
\end{figure} 

\newpage 

\begin{figure}

\begin{tabular}{ll} 
{\Large (a)}  & {\Large (b)} \\
\epsfxsize= 8.0cm\epsffile{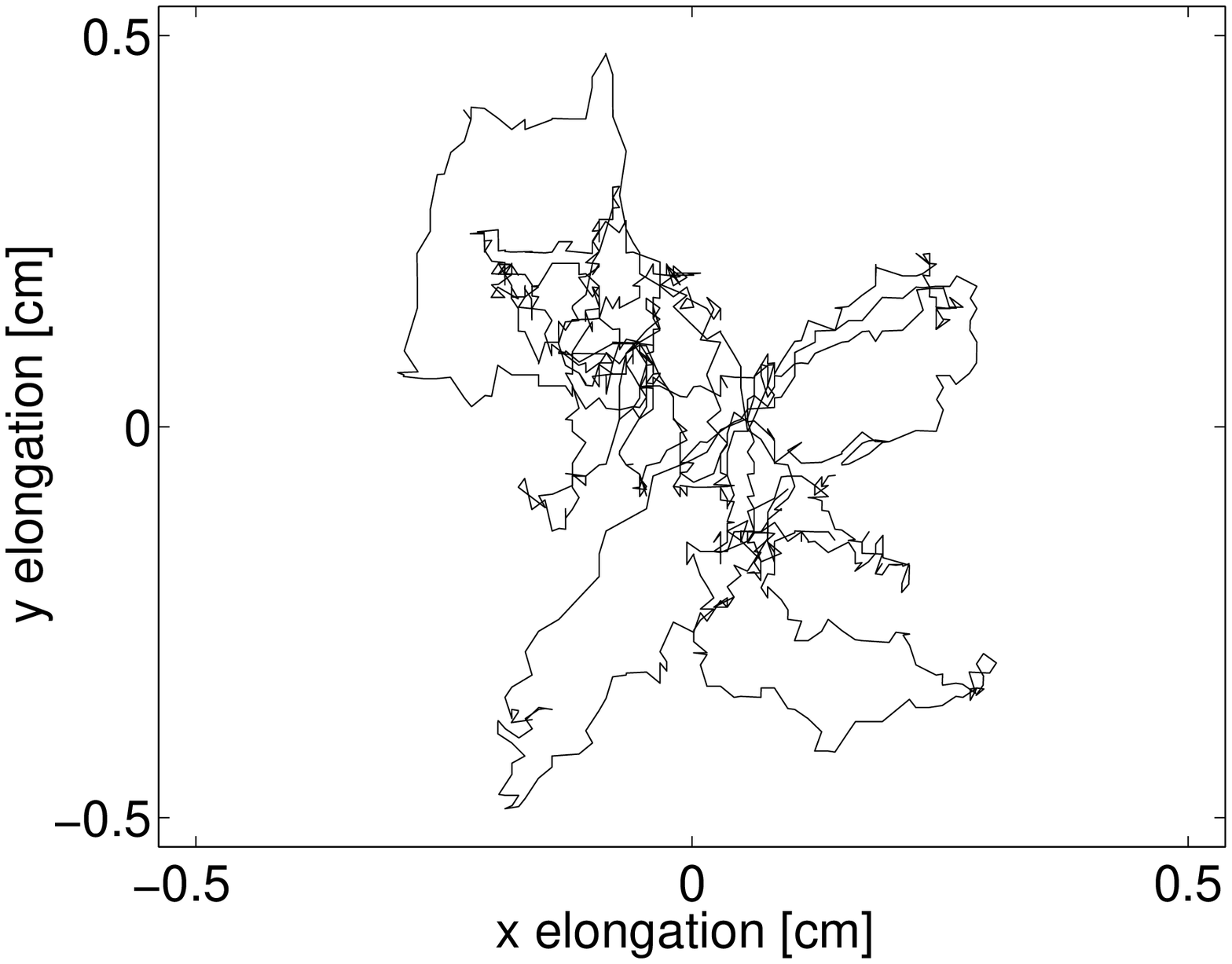} & 
\epsfxsize= 8.0cm\epsffile{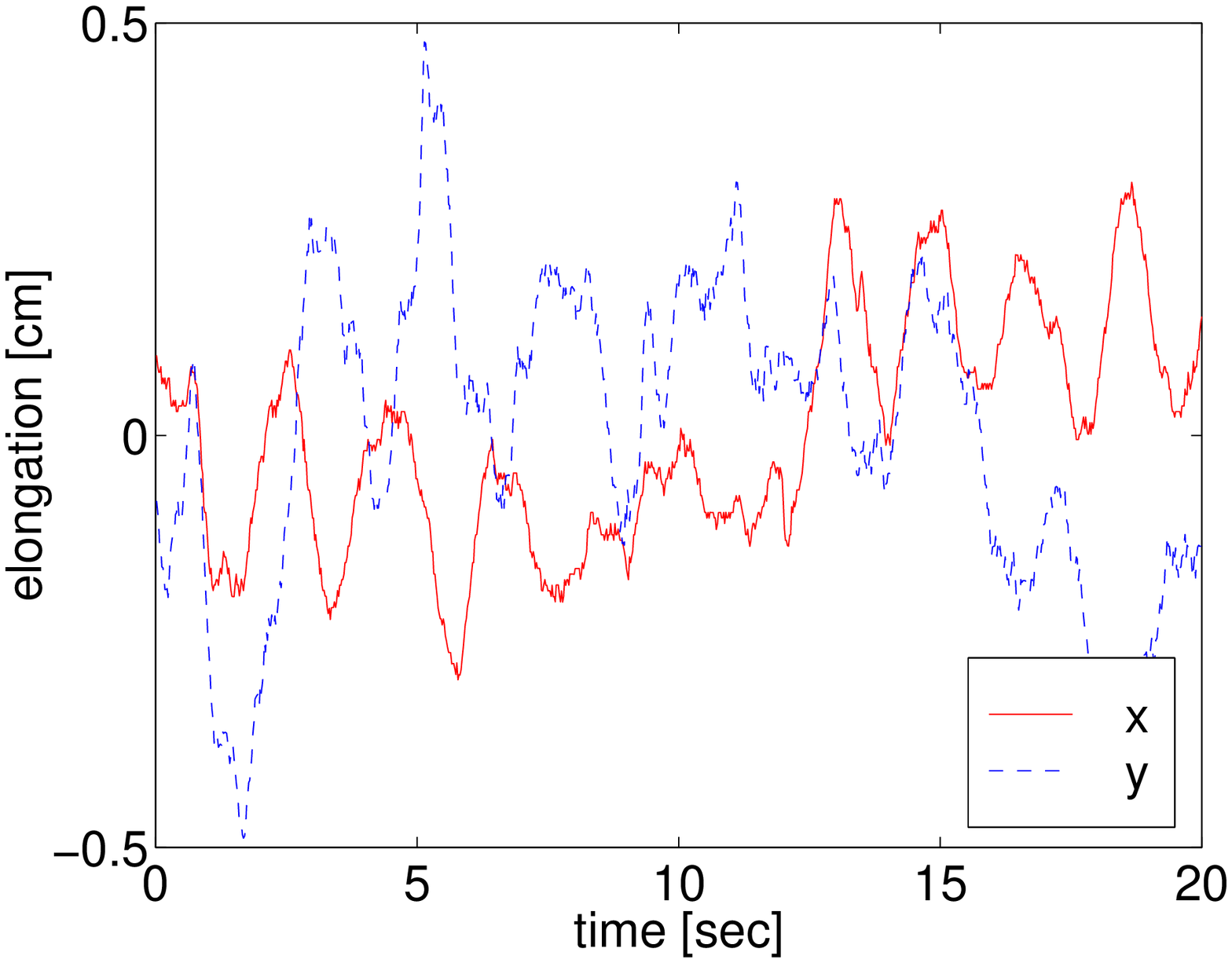} \\  
{\Large (c)}  & {\Large (d)} \\
\epsfxsize= 8.0cm\epsffile{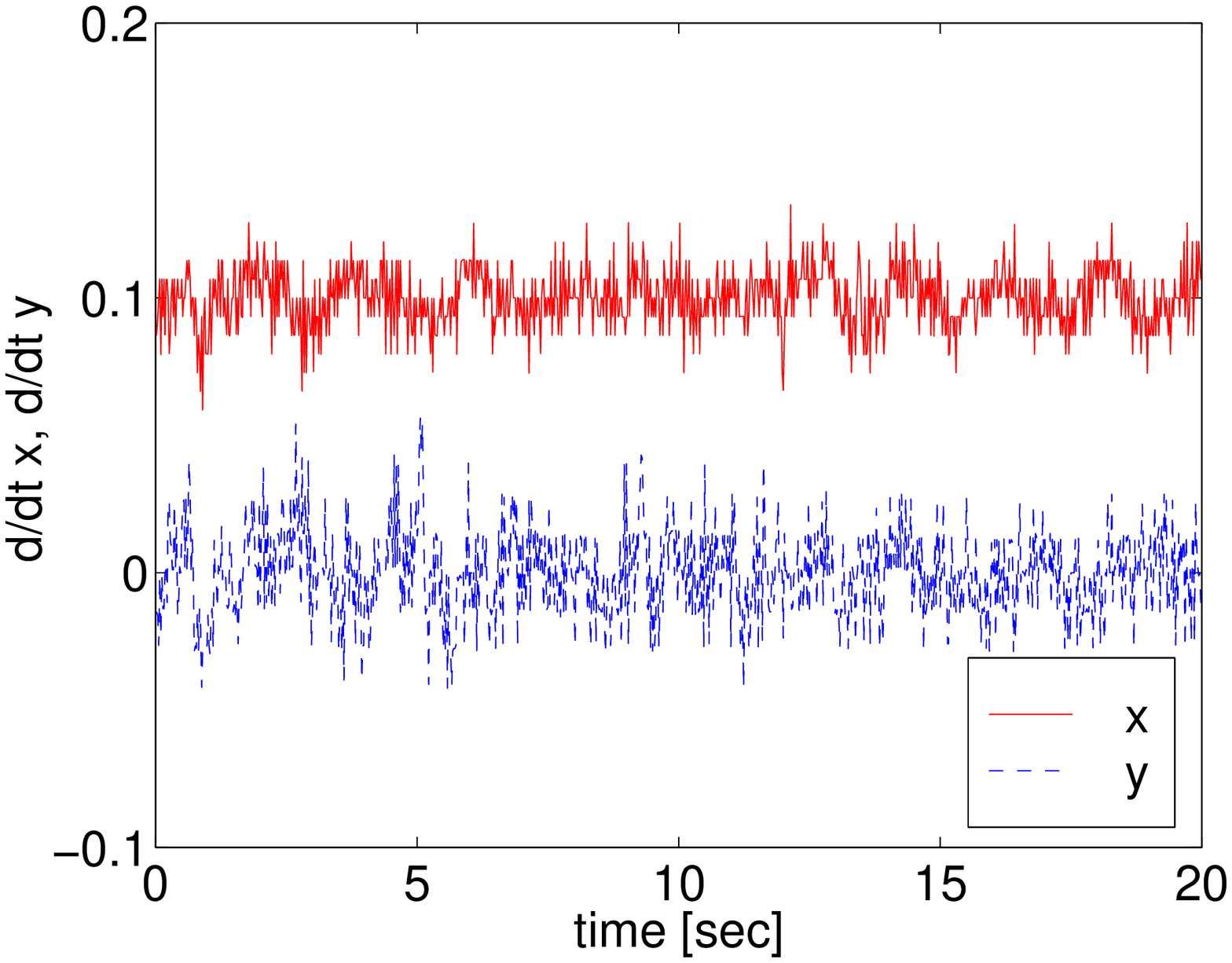}  & 
\epsfxsize= 8.0cm\epsffile{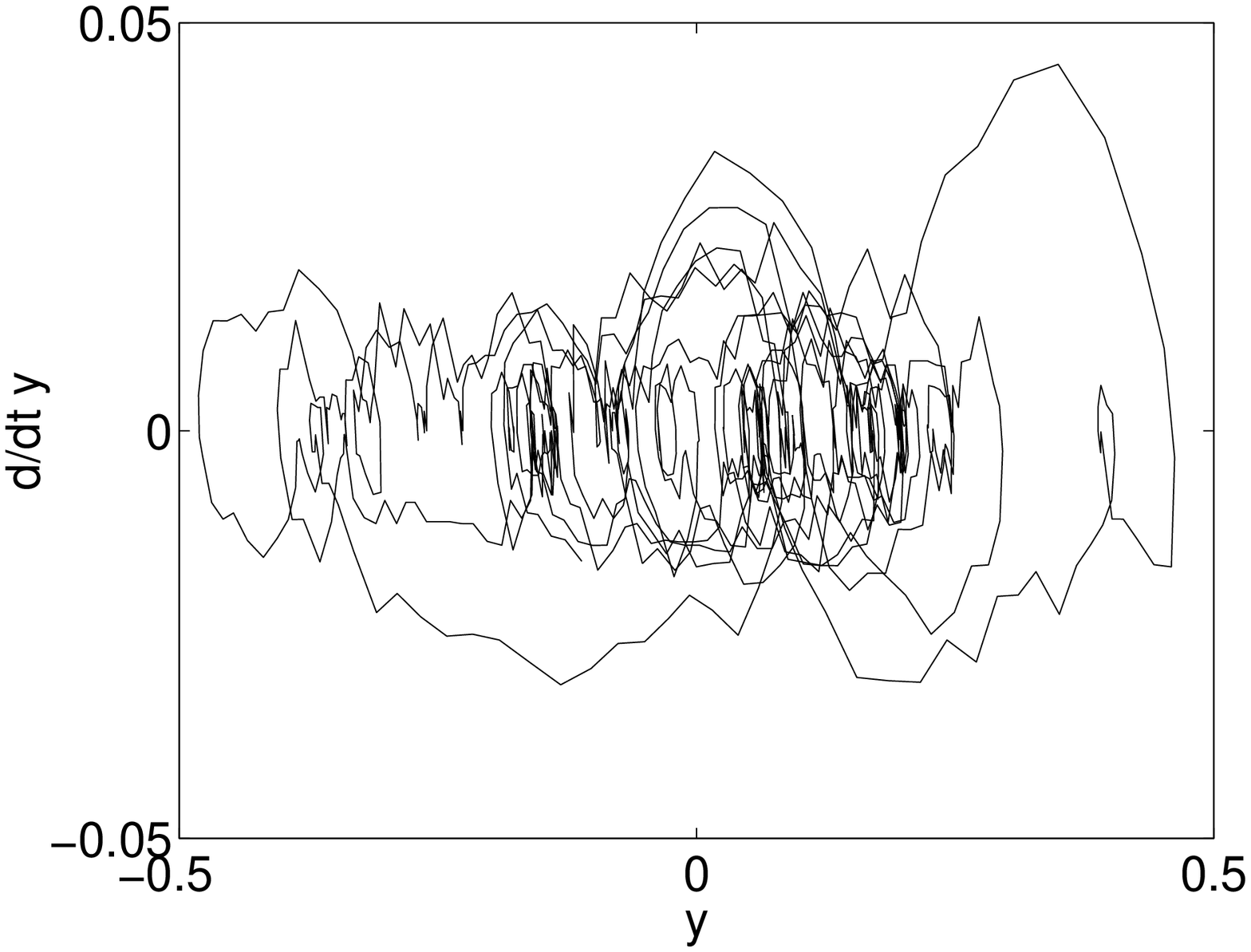} \\
\end{tabular}

\end{figure} 

\vfill 

\begin{center} 
{\Large FIG. 1} 
\end{center} 

\newpage 

\begin{figure}
\hspace{4cm} {\Huge eo }  \hspace{6cm} {\Huge  ecmp } \\
\epsfxsize=16.0cm\epsffile{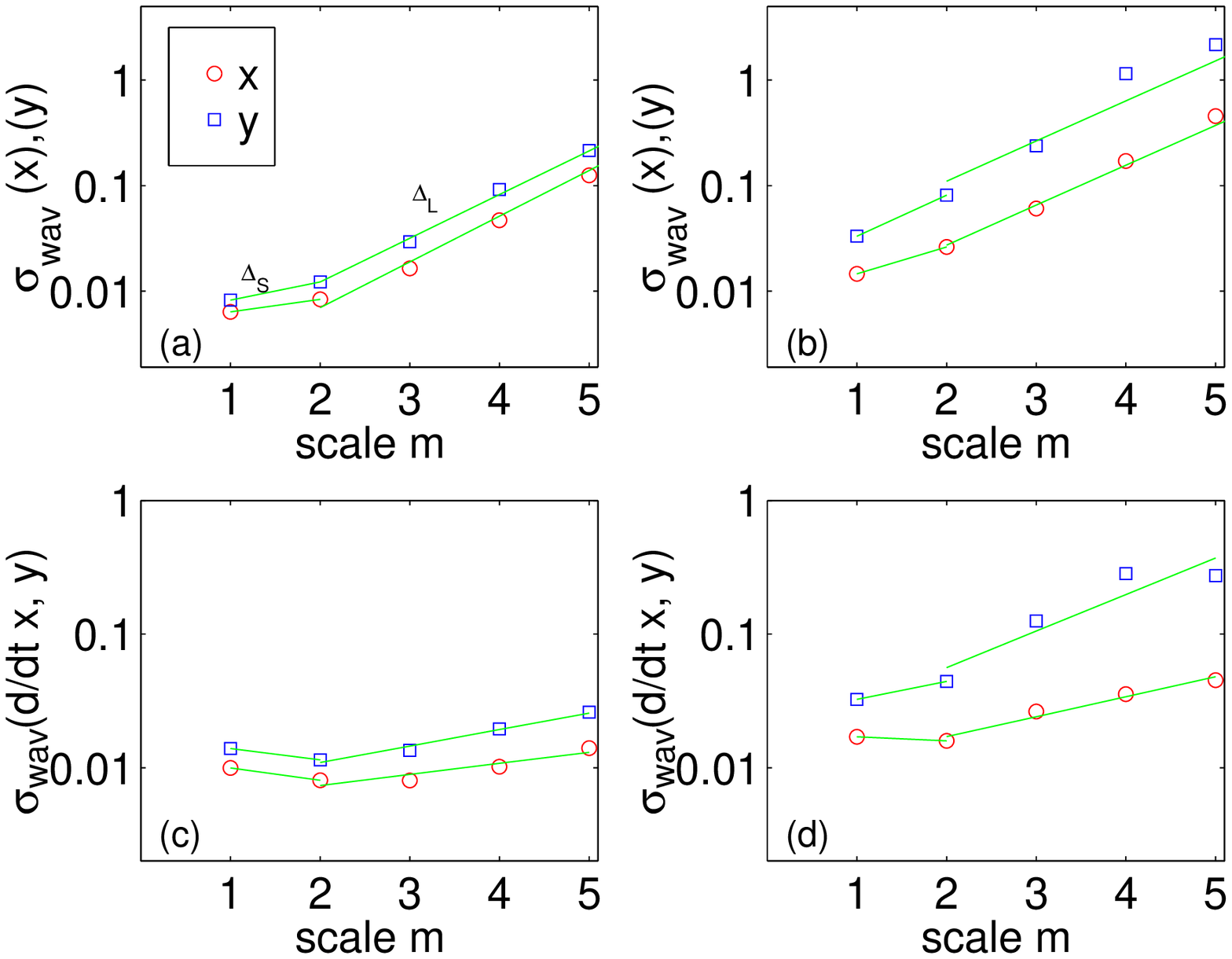} 
\end{figure} 

\vfill 

\begin{center} 
{\Large FIG. 2} 
\end{center} 

\newpage 

\begin{figure}
\begin{center} 
\noindent 
\epsfxsize= 15.0cm\epsffile{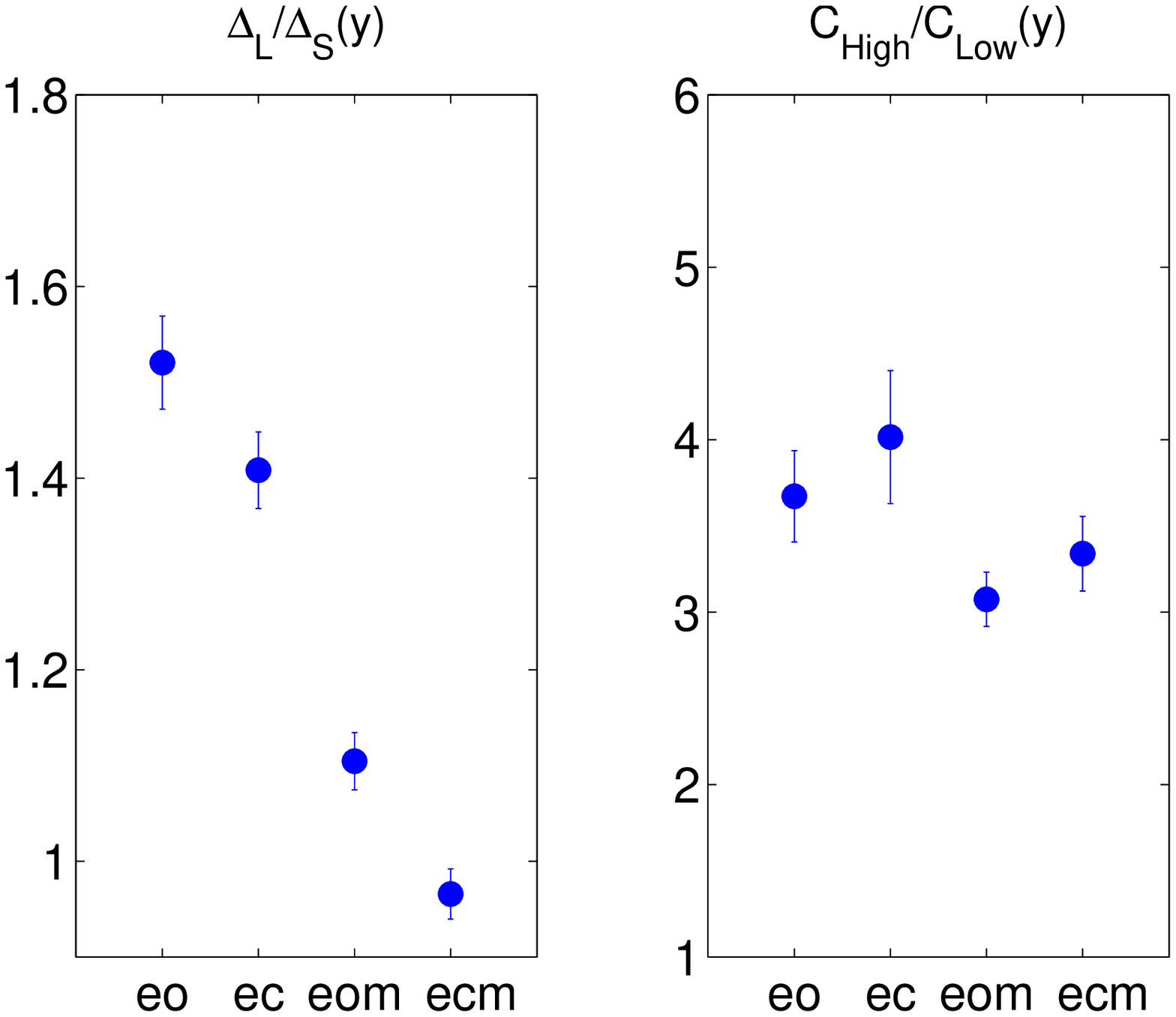} \\
\hspace{0.7cm} {\Large (a)} \hspace{6.5cm} {\Large (b)}  \\
\end{center} 
\end{figure} 

\vfill 

\begin{center} 
{\Large FIG. 3} 
\end{center} 

\newpage

\begin{figure}
\begin{center} 
\begin{tabular}{c}
\noindent 
\hspace{0.5cm} {\Large (a)} \\
\epsfxsize= 11.0cm\epsffile{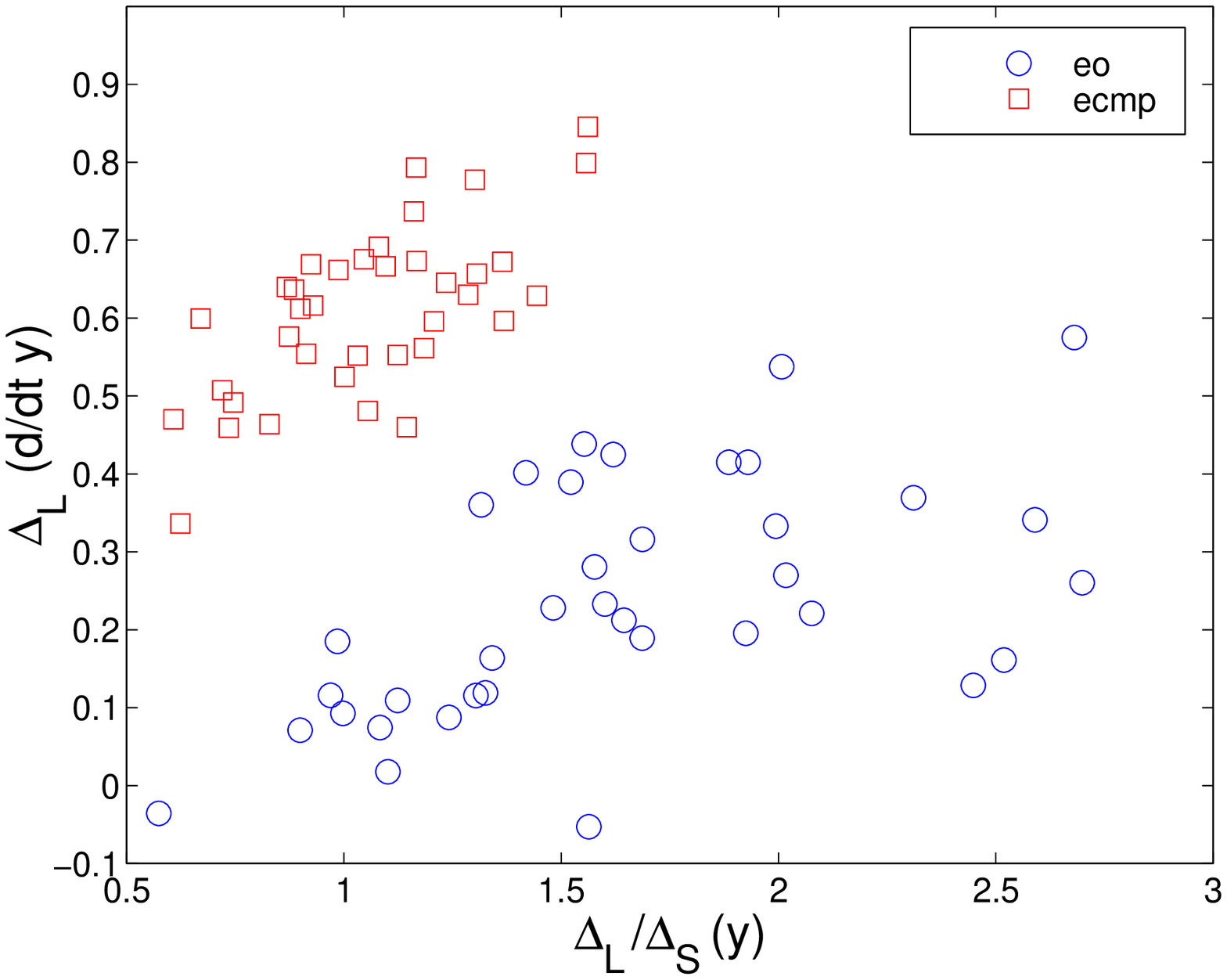} \\
{ \hspace{0.6cm} \Large (b) \hspace{4.4cm} (c) } \\
\epsfxsize= 11.0cm\epsffile{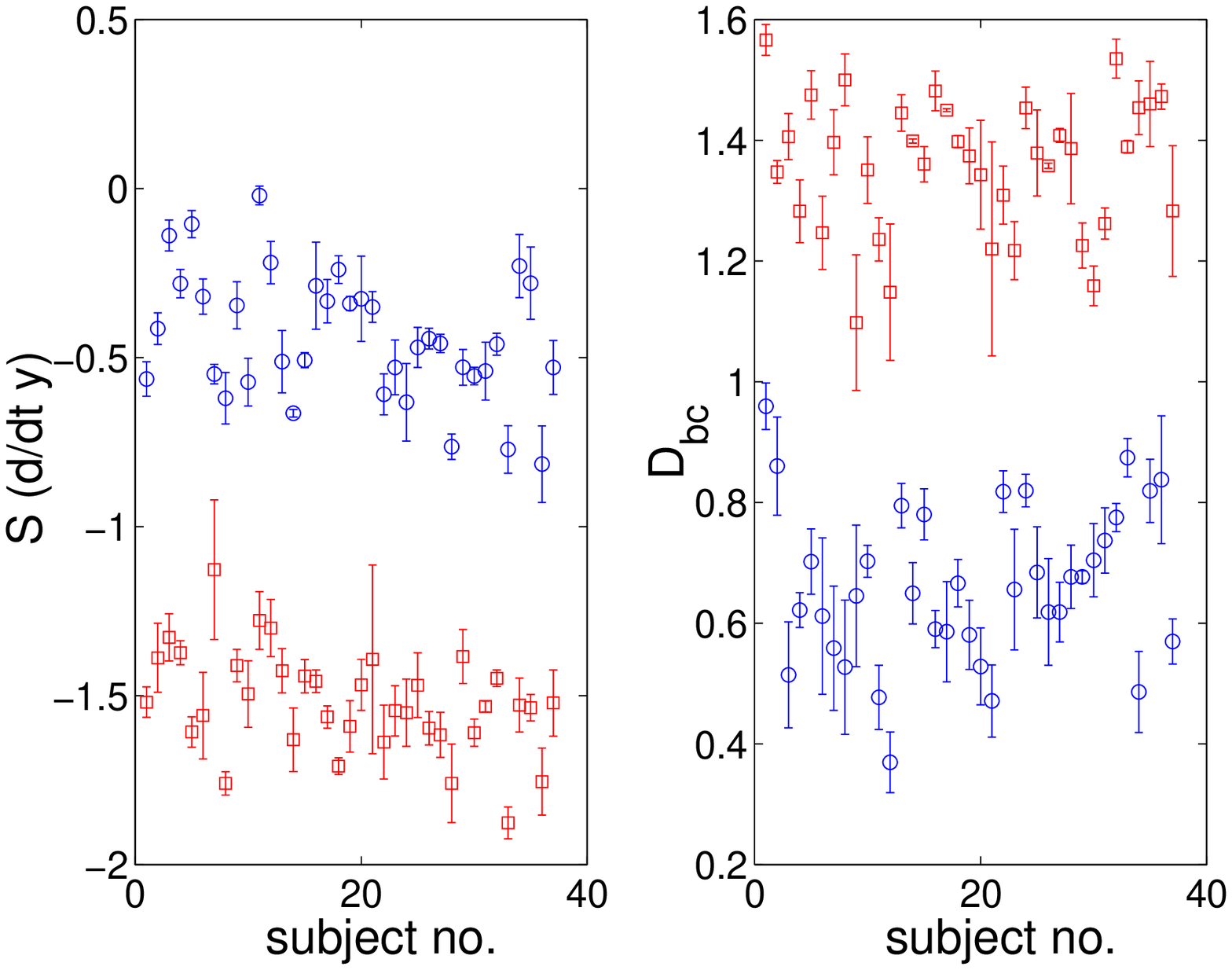} 
\end{tabular}
\end{center} 
\end{figure} 

\vfill 

\begin{center} 
{\Large FIG. 4} 
\end{center}


\begin{thebibliography}{99}


\bibitem{BAS94}   
J.B. Bassingthwaighte, L.S. Liebovitch and  B.J.  West,   
{ Fractal Physiology} 
(Oxford Univ. Press, New York, 1994).
%
\bibitem{THU981}   
S. Thurner, M.C. Feurstein and M.C.  Teich,  
{ Phys. Rev. Lett.} {\bf 80}, 1544-1547 (1998).
%
\bibitem{THU982}   
S. Thurner, M.C. Feurstein, S.B.  Lowen and M.C. Teich, 
{ Phys. Rev. Lett.} {\bf 81}, 5688-5691 (1998).
%
\bibitem{NEW93}
K.M. Newell, R.E.A.  Van Emmerik, D.  Lee and R.L. Sprague, 
{ Gait and Posture} {\bf 4}, 225-230 (1993).
%
\bibitem{MASS94}
J. Massion, 
{ Current Opinion in Neurobiology} {\bf 4}, 877-887 (1994).
%
\bibitem{KOHEN1} 
U. Oppenheim, R.  Kohen-Raz,  D. Alex, A. Kohen-Raz and  
A. Azarya,  
{ Diabetes Care} {\bf 22}, 328-332 (1999).  
%
\bibitem{KOHEN2} R. Kohen-Raz,  private communication.
%
\bibitem{JJC94}
J.J. Collins and C.J. DeLuca, 
Phys. Rev. Lett. {\bf 73} 764-767 (1994). 

\bibitem{JJC95}
C.C. Chow and J.J. Collins, 
Phys.  Rev. E {\bf 52} 907-912 (1995). 
%
\bibitem{steve}
 S.B.  Lowen and M.C.  Teich,  
{ Fractals} {\bf 3}, 183-210 (1995).
%
\bibitem{THU973}
S. Thurner, S.B.  Lowen, M.C. Feurstein, C.  Heneghan,  
H.G. Feichtinger  and M.C.  Teich,  
{ Fractals} {\bf 5}, 565-595 (1997).
%
\bibitem{scal} M.F. Shlesinger and B.J. West, 
Phys. Rev. Lett {\bf 67}, 2106-2108 (1991). 
%
\bibitem{JJC93}
J.J. Collins and  C.J.  De Luca,  
{ Exp. Brain Res.} {\bf 95}, 308-318 (1993). 
%
\bibitem{DOBE92}
I. Daubechies, 
{Ten Lectures on Wavelets}
(Society for Industrial and Applied Mathematics,  Philadelphia, PA,  1992).
%
\bibitem{MALL89}
S. Mallat,  
Trans. Amer. Math. Soc.  {\bf 315},  69-88 (1989).
%
\bibitem{MEYE86}
Y. Meyer,  
``Ondelettes, fonctions splines et analyses gradu\'{e}es,''
Lectures given at the University of Torino, Italy (1986).
%
\bibitem{ALDR96}
A. Aldroubi and M. Unser eds., 
{ Wavelets in Medicine and Biology}
(CRC Press, Boca Raton, FL, 1996).
%
\bibitem{ARNE95}
A. Arneodo, G.  Grasseau and M.   Holschneider,   
Phys. Rev. Lett. {\bf 61}, 2281-2284 (1988).
%
\bibitem{ABRY96}
P. Abry and P.  Flandrin,  in 
{ Wavelets in Medicine and Biology}
(CRC Press, Boca Raton, FL, 1996), pp. 413-437.
%
\bibitem{TEIC96}
M.C. Teich, C.  Heneghan, S.B.  Lowen and R.G.  Turcott, in
{ Wavelets in Medicine and Biology}
(CRC Press, Boca Raton, FL, 1996), pp. 383-412.
%
\bibitem{GOLD2}
A. Marrone, A.D. Polosa, G. Scioscia, S. Stramaglia, and A. Zenzola, 
{ Phys. Rev. E.} {\bf 60}, 1088-1091 (1999).
%
\bibitem{data} 
Data was gathered  with a ``Pro Balance Master'' (Neurocom Intl. Inc.).  

%
\end{thebibliography}
\end{document}